\documentclass[10pt]{article}

\usepackage{graphicx}
\usepackage{color}

\begin{document}

\newcommand{\ba}{\begin{eqnarray}}
\newcommand{\ea}{\end{eqnarray}}
\newcommand{\nn}{\nonumber}
\renewcommand{\d}{\textrm{d}}
\newcommand{\e}{\textrm{e}}
\renewcommand{\i}{\textrm{i}}
\newcommand{\p}{\partial}
\newcommand{\eps}{\epsilon}
\newcommand{\ve}{\varepsilon}
\newcommand{\cL}{\cal L}
\newcommand{\fr}{\tfrac}
\newcommand{\tr}{\textrm{tr}}

\renewcommand{\thefootnote}{\alph{footnote}}

\title{A model of glueballs} \author{Roman V. Buniy and Thomas
W. Kephart\\ \emph{Department of Physics and Astronomy}\\
\emph{Vanderbilt University, Nashville, TN 37235}}
%\email{roman.buniy@vanderbilt.edu}
%\author{Thomas W. Kephart}
%\email{kephartt@ctrvax.vanderbilt.edu}
%\affiliation{Department of Physics and Astronomy,\\
%Vanderbilt University, Nashville, TN 37325.}
\date{}
\maketitle

\begin{abstract}
We model the observed glueball mass spectrum in terms of energies for
tightly knotted and linked QCD flux tubes. The data is fit well with
one parameter. We predict additional glueball masses.
\end{abstract}

%\pacs{}

%\section{Introduction}

\emph{Introduction.---} The interpretation of non-$q\bar{q}$ states is
a puzzle with a long and controversial history~\cite{Swanson}. Many
experiments~\cite{PDG} report states that do not fit neatly into the
quark model. These states can be broadly classified as: (1)\ hybrids,
which are bound states of quarks and gluons like $q\bar{q}G$ with
quantum numbers $J^{PC}=0^{-+}$, $1^{-+}$, $1^{--}$, $2^{-+}$,
$\ldots$; (2)\ exotics, for example, four and six quark states such as
$qq\bar{q}\bar{q}$ and $qqq\bar{q}\bar{q}\bar{q}$ with quantum numbers
$J^{PC}=0^{--}$, $0^{+-}$, $1^{-+}$, $2^{+-},\ldots$; (3)\ glueballs
with pointlike or collective (e.g., strings \`{a} la
Nielsen--Olesen~\cite{Nielsen-Olesen}, or flux tubes) glue. Glueballs
do not contain valence quarks, but there could be sea/virtual quarks
within the glueball or in the currents that support the flux
tubes. From a bag model perspective one is led to suppose that the
lightest non--$q\bar{q}$ states are those with no constituent quarks,
i.e., the glueballs. Lattice calculations, QCD sum rules, electric
flux tube models, and constituent glue models leads to a consensus
that the lightest non--$q\bar{q}$ states are glueballs with quantum
numbers $J^{++}=0^{++}\ $and 2$^{++}$~\cite{West}. We will model all
$J^{++}$ states (i.e., all $f_{J}$ and $f'_J$ states listed by the
Particle Data Group (PDG)~\cite{PDG}), some of which will be
identified with rotational excitations, as knotted/linked
chromoelectric QCD flux tubes~\cite{comment1}.

Besides the fact they do not fit in the quark model~\cite{commentfit},
glueballs have some other characteristic signatures, including:
enhanced central production in gluon rich channels, branching
fractions incompatible with $q\bar{q}$ decay, reduced $\gamma\gamma$
coupling, and OZI suppresion. All the $J^{++}$ states we consider have
some or all of these properties. For instance, none have substantial
branching fractions to $\gamma\gamma$. However, mixing with $q\bar{q}$
isoscalar states can obscure some of the properties. A number of
candidates with masses below $2.5\,GeV$ have been identified. Beyond
their masses and widths, and some of their branching
ratios~\cite{PDG}, much remains to be learned about these states.

Knotted magnetic fields (which we will treat as solitons) have been
suggested as candidates for a number of plasma phenomena in systems
ranging from astrophysical, to atmospheric~\cite{lightning}, to
Bose-Einstein condensates~\cite{Bose-Einstein}. The energies of these
solitons are sometimes difficult to quantify since they depend on
parameters of the plasma, including temperature, pressure, density,
ionic content, etc.; however, we will argue that in QCD a well defined
soliton energy can be identified.

As has been shown in plasma physics, tight knots and links (defined
below) correspond to metastable minimum energy configurations. We will
argue by analogy that quantized tightly knoted and linked QCD flux
tubes are glueballs. (In what follows, we often use the term ``knots''
to mean knots and/or links.)

Movement of fluids often exhibits topological properties (for a
mathematical review see e.g.~\cite{Arnold}). For conductive fluids,
interrelation between fluid motion and magnetic fields via
magnetohydrodynamics may cause magnetic fields, in their turn, to
exhibit topological properties. For example, for a perfectly
conducting fluid, the (Abelian) magnetic helicity $L_H=\int\d ^3
x\,\eps_{ijk}A_i\p_jA_k$ is an invariant of motion~\cite{Woltier}, and
this quantity can be interpreted in terms of knottedness of magnetic
flux lines~\cite{Moffatt:1969}.

The dynamics of the magnetic fields follows the dynamics of the liquid
(magnetic flux lines are ``frozen'' into the fluid), and one finds
that a perfectly conducting, viscous and incompressible fluid relaxes
to a state of magnetic equilibrium without a change in
topology~\cite{Moffatt:1985}. As a result, for topologically
non-trivial plasma flows (with knotted streamlines), the ``freezing''
condition forces topological restrictions on possible changes in field
configurations. For linked non-intersecting loops $C_a$ with magnetic
fluxes $\Phi_a$, the helicity becomes~\cite{Moffatt:1969}
$L_H=\frac{k}{8\pi}\sum_{a\not=b}L(C_a,C_b)\Phi_a\Phi_b$. Here
$L(C_a,C_b)$ is the Gauss linking number~\cite{helicity_comment}. By
its topological nature, helicity in the QCD flux can be one of the
quantum numbers characterizing glueballs. However, there is another
invariant called the knot energy that is less obvious but as important
in the classification of solitonic knots.

%\section{Knot energies}

\emph{Knot energies.---} Consider a hadronic collision that produces
some number of baryons and mesons plus a gluonic state in the form of
a closed QCD flux tube (or a set of tubes). From an initial state, the
fields in the flux tubes quickly relax to an equilibrium
configuration, which is topologically equivalent to the initial
state. (We assume topological quantum numbers are conserved during
this rapid process.) The relaxation proceeds through minimization of
the field energy. Flux conservation and energy minimization force the
fields to be homogeneous across the tube cross sections. This process
occurs via shrinking the tube length, and halts to form a ``tight''
knot or link. The radial scale will be set by $\Lambda
_{\textrm{QCD}}^{-1}$. The energy of the final state depends only on
the topology of the initial state and can be estimated as follows. An
arbitrarily knotted tube of radius $a$ and length $l$ has the volume
$\pi a^2 l$. Using conservation of flux $\Phi_E$, the energy becomes
$\propto l(\tr\Phi_E^2)/(\pi a^2)$. Fixing the radius of the tube (to
be proportional to $\Lambda_{\textrm{QCD}}^{-1}$), we find that the
energy is proportional to the length $l$. The dimensionless ratio
$\ve(K)=l/(2a)$ is a topological invariant and the simplest definition
of the ``knot energy''~\cite{knot_energy}.

Many knot energies have been calculated by Monte Carlo
methods~\cite{knot-energies} and certain types can be calculated
exactly (see below), while for other cases simple estimates can be
made (see Table~\ref{table}). For example, the knot energy of the
connected product of two knots $K_1$ and $K_2$ satisfies \ba
\ve(K_1\#K_2)<\ve(K_1)+\ve(K_2)\ea A rule of thumb is \ba
\ve(K_1\#K_2)\approx \ve(K_1)+\ve(K_2)-(2\pi-4),\ea which results from
removing two half tori, one from each knot, and replacing these with
two connecting cylinders of lengths $a$. This, for example, agrees
with the Monte Carlo values for $\ve(3_1\#3_1)$ and $\ve(3_1\#3_1^*)$
to about $5\%$.

Most of the knot energies in Table 1 have been taken from
\cite{knot-energies}, but we have independently calculated the energy
of $2^2_{1}$, $4^3_1$ and $6^4_1$ exactly and the energy for several
other knots and links approximately. We find $\ve(2^2_{1})=4\pi\approx
12.57$, to be compared with the Monte Carlo value $12.6$. We also find
$\ve(4^3_{1})=6\pi+2$ and $\ve(6^4_{1})=8\pi+3$, where there are no
Monte Carlo comparisons available, or needed.

%\section{Model}

\emph{Model.---} In our model, the chromoelectric
fields~\cite{comment3} $F_{0i}$ are confined to knotted tubes, each
carrying one quantum of conserved flux~\cite{flux}~\cite{soliton}. We
consider a stationary Lagrangian density \ba{\cL}=\frac{1}{2}\tr
F_{0i}F^{0i}-V,\label{L}\ea where, similar to the MIT bag
model~\cite{MIT-bag}, we included the possibility of a constant energy
density $V$. To account for conservation of the flux $\Phi_E$, we add
to (\ref{L}) the term \ba\tr\lambda\{\Phi_E/(\pi
a^2)-n^iF_{0i}\},\label{flux1}\ea where $n^i$ is the normal vector to
a section of the tube and $\lambda$ is a Lagrange multiplier. Varying
the full Lagrangian with respect to $A_\mu$, we find \ba
D^0(F_{0i}-\lambda n_i)=0,\ \ \ \ \ D^i(F_{0i}-\lambda
n_i)=0,\label{flux2}\ea which has the constant field \ba
F_{0i}=(\Phi_E/\pi a^2)n_i\label{F}\ea as its solution. With this
solution, the energy is positive and proportional to $l$ and thus the
minimum of the energy is achieved by shortening $l$, i.e. tightening
the knot.

\begin{figure*}
\centering \includegraphics[angle=0]{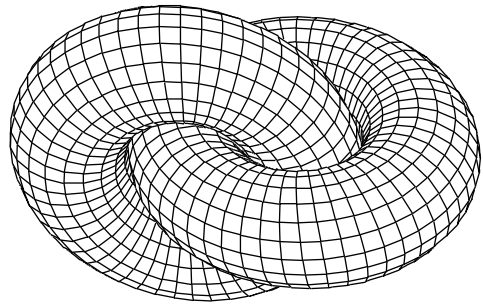}
\caption{\label{figure1} The shortest knot/link solitonic flux
configuration has the topology of two linked tori, which in knot
theory notation is $2^2_1$. This corresponds to the lightest glueball
candidate $f_0(600)$.}
\end{figure*}

\begin{figure*}
\centering \includegraphics[angle=0]{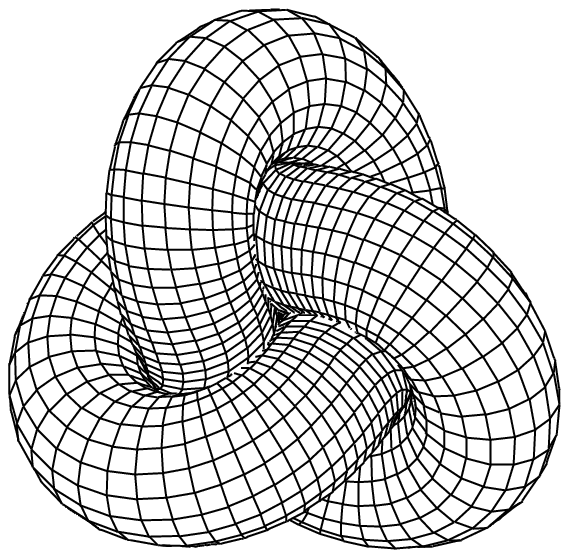}
\caption{\label{figure2} The second shortest solitonic flux
configuration is the trefoil knot $3_1$ corresponding to the second
lightest glueball candidate $f_0(980)$.}
\end{figure*}

We proceed to identify knotted and linked QCD flux tubes with
glueballs, where we include all $f_J$ and $f'_J$ states. The lightest
candidate is the $f_0(600)$, which we identify with the shortest
knot/link, i.e., the $2^2_1$ link (see Figure~\ref{figure1}); the
$f_0(980)$ is identified with the next shortest knot, the $3_{1}$
trefoil knot (see Figure~\ref{figure2}), and so forth. All knot and
link energies have been calculated for states with energies less then
$1680\,\textrm{MeV}$. Above $1680\,\textrm{MeV}$ the number of knots
and links grows rapidly, and few of their energies have been
calculated. However, we do find knot energies corresponding to known
$f_J$ and $f'_J$ states, and so can make preliminary identifications
in this region. (We focus on $f_J$ and $f'_J$ states from the PDG
summary tables. The experimental errors are also quoted from the
PDG. There are a number of additional states reported in the extended
tables, but some of this data is either conflicting or inconclusive.)

Our detailed results are collected in Table \ref{table}, where we list
$f_J$ and $f'_J$ masses, widths, and our identifications of these
states with knots, together with the corresponding knot energies.

\begin{figure*}
\centering \includegraphics[angle=0]{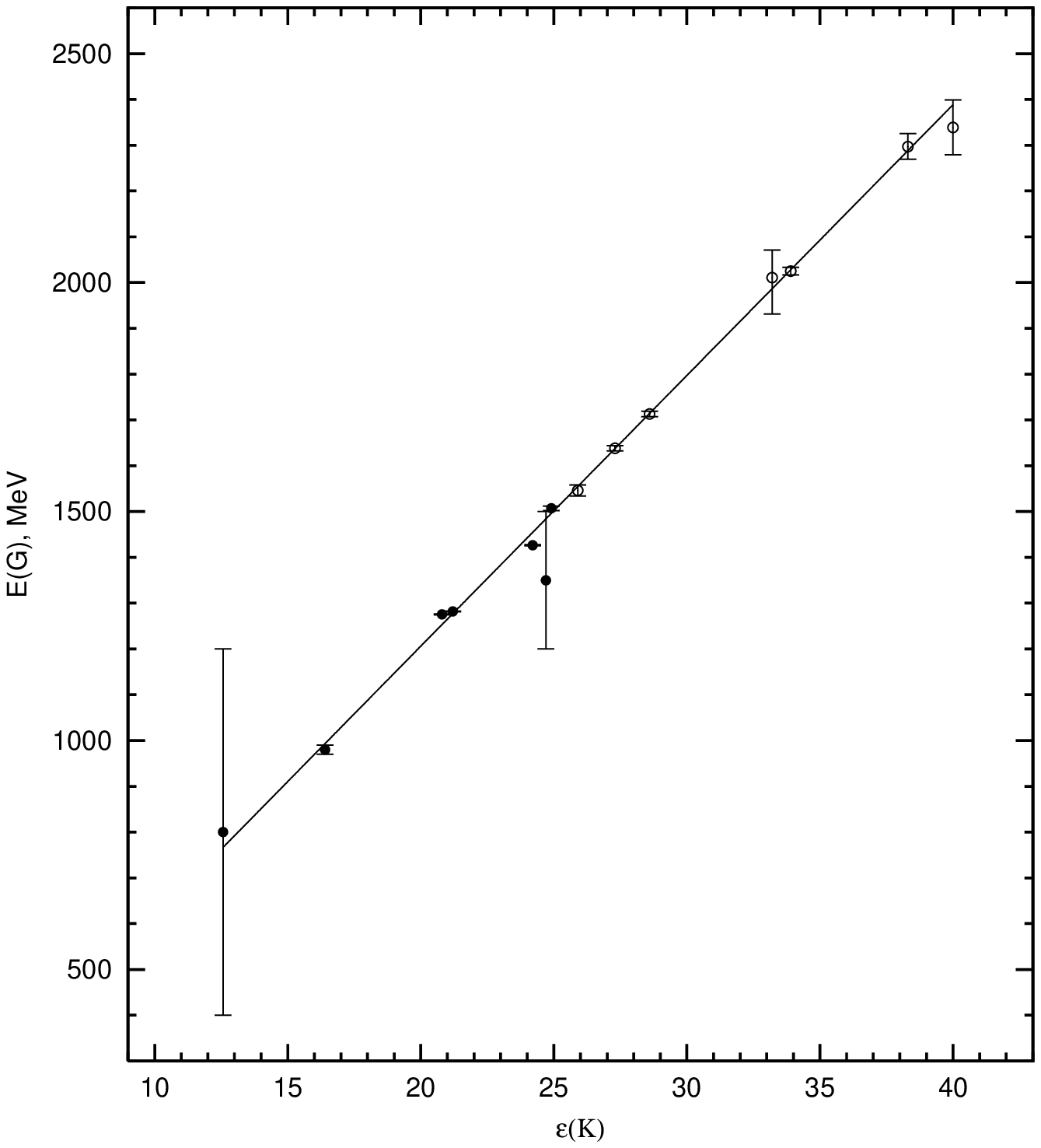}
\caption{\label{figure}Relationship between the glueball spectrum
$E(G)$ and knot energies $\ve(K)$. Each point in this figure
represents a glueball identified with a knot or link. The straight
line is our model and is drawn for the fit (\ref{fit1}).}
\end{figure*}

In Figure~\ref{figure} we compare the mass spectrum of $f$ states with
the identified knot and link energies. Since errors for the knot
energies in~\cite{knot-energies} were not reported, we conservatively
assumed the error to be $1\%$. A least squares fit to the most
reliable data (below $1680\,\textrm{MeV}$) gives \ba E(G)=(23.4\pm
46.1)+(59.1\pm 2.1)\ve(K)\ \ \ [\textrm{MeV}],\label{fit1}\ea with
$\chi^2=9.1$. The data used in this fit is the first seven $f_J$
states (filled circles in Figure~\ref{figure}) in the PDG summary
tables. Inclusion of the remaining seven (non-excitation) states
(unfilled circles in Figure~\ref{figure}) in Table~\ref{table}, where
either the glueball or knot energies are less reliable, does not
significantly alter the fit and leads to \ba E(G)=(26.9\pm
24.9)+(58.9\pm 1.0)\ve(K)\ \ \ [\textrm{MeV}],\label{fit2}\ea with
$\chi^2=10.1$. The fit (\ref{fit1}) is in good
agreement~\cite{comment2} with our model, where $E(G)$ is proportional
to $\ve(K)$. Better HEP data and the calculation of more knot energies
will provide further tests of the model and improve the high mass
identification.

In terms of the bag model~\cite{MIT-bag}, the interior of tight knots
correspond to the interior of the bag. The flux through the knot is
supported by current sheets on the bag boundary (surface of the
tube). Knot complexity can be reduced (or increased) by unknotting
(knotting) operations~\cite{Rolfsen,Kauffman}. In terms of flux tubes,
these moves are equivalent to reconnection
events~\cite{reconnection}. Hence, a metastable glueball may decay via
reconnection. Once all topological charge is lost, metastability is
lost, and the decay proceeds to completion. Two other glueball decay
processes are: flux tube (string) breaking; this favors large decay
widths for configurations with long flux tube components; and quantum
fluctuations that unlink flux tubes; this would tend to broaden states
with short flux tube components. As yet we are not able to go beyond
these qualitative observations, but hope to be able to do so in the
future.

We have assumed one fluxoid per tube. There may be states with more
than one fluxoid, but these would presumably have somewhat fatter flux
tubes with higher flux densities and higher energies. For example, the
two fluxoid trefoil knot $3_1$ would certainly have
$\ve(K)>2\,\ve(3_1)$ and a fairly reliable estimate gives
$\ve(K)\approx 2\sqrt{2}\,\ve(3_1)$. Hence most multifluxoid states
would be above the mass range of known glueballs.

%\section{Discussions and conclusions}

\emph{Discussions and conclusions.---} In principle, lattice
calculations can find any tame knot (knot without an infinite number
of crossings or other pathology \cite{Rolfsen}) configuration, since
there is always a contour through the lattice that represents the
knotted path by some specific Wilson loop. However, since one is
constrained by the rigidity of the lattice, energy minimization is
difficult and requires a very fine-grained lattice.  Thus we expect
shape-evolving Monte Carlo techniques \cite{knot-energies} to be much
more efficient and accurate for this purpose.

Now we must discuss the details of identifications made in
Table~\ref{table}. First, the $f_2(1270)$ does have a quark model
interpretation. Either the glueball state in this range is well mixed
with the quark model state and is part of the resonance at
$1275\,\textrm{MeV}$ (our interpretation), or the glueball state is
yet to be discovered. In either case, more data in this region would
be helpful. Next, the four (unconfirmed) glueball states with masses
less then $1680\,\textrm{MeV}$ from the extended PDG tables are
identified as follows: (1) the $4^2_1$ link with $E(G)=1289$ and the
$4_1$ knot with $E(G)=1277$ are nearly degenerate, and the $f_1(1285)$
could actually be a pair of nearly degenerate states with identical
quantum numbers associated with these knots; this is a possible
interpretation of the $f_1(1285)$ mass measurements summarized on page
481 of Ref.~\cite{PDG}; (2) the $f_2(1430)$ is treated as a rotational
excitation of the $f_1(1420)$ and identified with the $5_1$ knot; the
energy difference between these two states, $\delta'$, is a few MeV,
but not well determined; this difference is of the order of what one
would expect for rotational excitations; [We approximate
$E(f_J)=E(f_0)+\frac{1}{2}J(J+1)\delta$.] (3) we treat the $f_1(1510)$
as the first and the $f'_2(1525)$ could be the second~\cite{f1525}
rotational excitation of the $f_0(1500)$, which we identify with the
$5_2$ knot; now the energy step size is $\delta\approx 5\textrm{MeV}$
which agrees with a simple estimate; (4) we assign the $f_2(1565)$ and
the $f_2(1640)$ to the $5^2_1$ and the $6^3_3$ links respectively.

Further details of knot excitations would be interesting to
investigate, as would quantum and curvature corrections. At present we
do not have a reliable way to estimate all these effects, nor do we
have a good way to calculate glueball decays. However, we do expect
high mass glueball production to be suppressed because more
complicated non-trivial topological field configurations are
statistically disfavored.

Finally, knot solitons may also be able to survive within a
quark-gluon plasma (e.g., in the interior of a RHIC event, quark star,
or in the early universe). Complications will certainly arise in these
cases due to additional parameters describing the media, as with
knotted and linked electromagnetic plasma solitons; but if one holds
the parameters constant throughout the region of interest in this or
any sytem that supports knot/link solitons, the energy spectrum will
be universal up to a scaling.

\emph{Acknowledgments.---} We thank Med Webster, Kevin Stenson, Eric
Vaandering, Will Johns, Tom Weiler, and Jack Ng for useful comments
and discussions. This work was supported by U.S. DoE grant number
DE-FG05-85ER40226.

%\bibliographystyle{unsrt}
%\bibliography{general,knots,mhd}

\begin{thebibliography}{99}
\bibitem{comment1} In terms of 't Hooft's dual confinement picture,
confinement is due to monopole condensation, with flux tubes being
chromoelectric.  However, since we as yet have no soliton solutions to
the full QCD theory, we are not married to any confinement scheme; all
we require are tightly knotted/linked flux tubes (chromoelectric,
chromomagnetic, or even chromodyonic) of uniform cross section, though
we prefer to go with conventional interpretation. One could
potentially gain information on confinement from an analysis of
glueball decay (see the discussion below).  Also, one could
potentially model hybrid states or even exotics as knotted/linked
color flux tube type bags with valance quarks inside. Although this
topic is outside the scope of the current discussion, we hope to
address it elsewhere.
\bibitem{commentfit} We note that attempts have been made to fit
several of the $f_J^{++}$ states into the quark model.
\bibitem{flux} In a discussion of flux quantization in non-Abelian
gauge theories, it is important to keep the following fact in mind.
For a static configuration, and using the temporal gauge, we can avoid
the creation of chromomagnetic flux lines inside the flux
tubes~\cite{Nielsen-Olesen}. The dual argument shows chromoelectric
flux creation can be avoided as well.
\bibitem{comment2} Note that $\chi ^{2}=251$ for a fit where the first
glueball is missed out, and $\chi ^{2}=355$ for a fit where the first
knot/link is missed out. This is strong evidence that our
identification is appropriate.
\bibitem{f1525} However, there is strong evidence that the
$f'_2(1525)$ fits well into the quark model. If we drop it from our
table, the $\chi^2$ is unaffected since we only fit the ground state
knots, not their excitations.
\end{thebibliography}

\newpage

\begin{center}
\begin{table}[htb]
\begin{center}
\caption{\label{table}Comparison between the glueball mass spectrum
and knot energies.}
\begin{tabular}{cccccc}\hline\hline
{\rule[-3mm]{0mm}{8mm} State} & Mass & Width & $K$~\footnotemark[1] &
$\ve(K)$~\footnotemark[2] & $E(G)$~\footnotemark[3] \\ \hline
{\rule[1mm]{0mm}{3mm} $f_0(600)$} & $400-1200$ & $600-1000$ & $2^2_1$
& $12.6\ [4\pi]$ & $768\ [766]$\\ $f_0(980)$ & $980\pm 10$ & $40-100$
& $3_1$ & $16.4$ & $993$\\ $f_2(1270)$ & $1275.4\pm 1.2$ &
$185.1^{+3.4}_{-2.6}$ & $2^2_1*0_1$ & $[6\pi+2]$ & $[1256]$\\
$f_1(1285)$ & $1281.9\pm 0.6$ & $24.0\pm 1.2$ & $4_1$ & $21.2$ &
$1277$\\ & & & $4^2_1$ & $(21.4)$ & $(1289)$\\ $f_1(1420)$ &
$1426.3\pm 1.1$ & $55.5\pm 2.9$ & $5_1$ & $24.2$ &
$1454$\\$\{f_2(1430)$ & $\approx 1430\}$~\footnotemark[4] & & $5_1$ &
$24.2$ & $1454+\delta'$\\ $f_0(1370)$ & $1200-1500$ & $200-500$ &
$3_1*0_1$ & $(24.7)$ & $(1484)$\\$f_0(1500)$ & $1507\pm 5$ & $109\pm
7$ & $5_2$ & $24.9$ & $1496$\\ $\{f_1(1510)$ & $1518\pm 5$ & $73\pm
25\}$ & $5_2$ & $24.9$ & $1496+\delta$\\$f'_2(1525)$ & $1525\pm 5$ &
$76\pm 10$ & $5_2$ & $24.9$ & $1496+3\delta$\\ $\{f_2(1565)$ &
$1546\pm 12$ & $126\pm 12\}$ & $5^2_1$ & $(25.9)$ & $(1555)$\\
$\{f_2(1640)$ & $1638\pm 6$ & $99^{+28}_{-24}\}$ & $6^3_3$ &
$((27.3))$ & $((1638))$\\ \multicolumn{6}{c}{.\dotfill.}\\ & & &
$(2^2_1*0_1)*0_1$~\footnotemark[5] & $[8\pi+3]$ &
$[1686]$~\footnotemark[6]\\$f_0(1710)$ & $1713\pm 6$ & $125\pm 10$ &
$6^3_2$ & $((28.6))$ & $((1714))$\\ & & & $3_1\#3_1^*$ & $28.9\
(30.5)$ & $1732\ (1827)$\\ & & & $3_1\#3_1$ & $29.1\ (30.5)$ & $1744\
(1827)$\\ & & & $2^2_1*2^2_1$ & $[8\pi+4]$ & $[1745]$\\ & & & $6_2$ &
$29.2$ & $1750$\\ & & & $6_1$ & $29.3$ & $1756$\\ & & & $6_3$ & $30.5$
& $1827$\\ & & & $7_1$ & $30.9$ & $1850$\\ & & & $8_{19}$ & $31.0$ &
$1856$\\ & & & $8_{20}$ & $32.7$ & $1957$\\ $f_2(2010)$ &
$2011^{+60}_{-80}$ & $202\pm 60$ & $7_2$ & $33.2$ & $1986$\\
$f_4(2050)$ & $2025\pm 8$ & $194\pm 13$ & $8_{21}$ & $33.9$ & $2028$\\
& & & $8_1$ & $37.0$ & $2211$\\ & & & $10_{161,162}$ & $37.6$ &
$2247$\\ $f_2(2300)$ & $2297\pm 28$ & $149\pm 40$ & $8_{18}$, $9_1$ &
$38.3$ & $2288$\\ $f_2(2340)$ & $2339\pm 60$ & $319^{+80}_{-70}$ &
$9_2$ & $40.0$ & $2389$\\ & & & $10_1$ & $44.8$ & $2672$\\
{\rule[-3mm]{0mm}{3mm} } & & & $11_1$ & $47.0$ & $2802$\\\hline\hline
\end{tabular}
\end{center}
\end{table}
\end{center}

\footnotetext[1]{Notation $n^l_k$ means a link of $l$ components with
$n$ crossings, and occurring in the standard table of links (see
e.g. \protect\cite{Rolfsen}) on the $k^\textrm{th}$ place. $K\#K'$
stands for the knot product (connected sum) of knots $K$ and $K'$ and
$K*K'$ is the link of the knots $K$ and $K'$.} \footnotetext[2]{Values
are from \protect\cite{knot-energies} except for our exact
calculations of $2^2_1$, $2^2_1*0_1$, and $(2^2_1*0_1)*0_1$ in square
brackets, our analytic estimates given in parentheses, and our rough
estimates given in double parentheses.}  \footnotetext[3]{$E(G)$ is
obtained from $\ve(K)$ using the fit (\ref{fit1}).}
\footnotetext[4]{States in braces are not in the PDG summary tables.}
\footnotetext[5]{This is the link product that is not $2^2_1*2^2_1$.}
\footnotetext[6]{Resonances have been seen in this region, but are
unconfirmed~\cite{PDG}.}

\end{document}